# "Inside Out" Growth Method for High-Quality Nitrogen-Doped Graphene

*Sara Fiori[a,b], Daniele Perilli[c], Mirco Panighel[b], Cinzia Cepek[b], Aldo Ugolotti[c], Alessandro Sala[a,b], Hongsheng Liu[c,1], Giovanni Comelli[a,b], Cristiana Di Valentin[c]\* and Cristina Africh[b]\**

[a] Department of Physics, University of Trieste, via A. Valerio 2, 34127, Trieste, Italy

[b] CNR-IOM, Laboratorio TASC, S.S. 14 Km 163.5, Basovizza, 34149, Trieste, Italy

[c] Dipartimento di Scienza dei Materiali, Università di Milano-Bicocca, via R. Cozzi 55, I-20125 Milano, Italy

[1] Laboratory of Materials Modification by Laser, Ion and Electron Beams, Dalian University of Technology, Ministry of Education, Dalian 116024, China

\*Corresponding author (experiments). E-mail: africh@iom.cnr.it (Dr. Cristina Africh)

\*Corresponding author (theory). E-mail: cristiana.divalentin@unimib.it (Prof. Cristiana Di Valentin)

**Abstract:**

High-quality nitrogen-doped graphene on nickel is prepared by exploiting both the catalytic properties of nickel and the solubility of nitrogen atoms into its bulk. Following the standard chemical vapor deposition procedure, a previously nitrogen-doped nickel substrate is exposed to carbon-containing precursors so that nitrogen atoms, segregating to the surface, remain trapped in the growing graphene network. Morphological and chemical characterization by scanning tunneling microscopy and X-ray photoelectron spectroscopy demonstrates that the process yields a flat, wide, continuous nitrogen-doped graphene layer. Experimental results are combined with a thorough density functional theory investigation of possible structural models, to obtain a clear description at the atomic scale of the various configurations of the nitrogen atoms observed in the graphene mesh. This growth method is potentially scalable and suitable for the production of high-performance nano-devices with well-defined nitrogen centers, to be exploited as metal-free carbon-based catalysts in several applicative fields such as electrochemistry, energy storage, gas storage/sensing or wastewater treatment.

## 1. Introduction

Graphene (Gr), the one-atom thick 2D material isolated in 2004 by Novoselov and Geim [1], is formed by $sp^2$-hybridized carbon (C) atoms, arranged in a honeycomb lattice. This structure is responsible for its peculiar properties, such as high electrical and thermal conductivity, charge mobility, transparency and flexibility, which are being successfully exploited in several research fields [2-6]. Nevertheless, pristine Gr suffers some severe limitations. Most notably, pristine Gr is a zero-band gap semiconductor, thus preventing its practical implementation in electronic devices.

Nowadays, doping of Gr is claimed to be a promising approach in order to tune its properties: for example, on the basis of *ab initio* calculations, it has been reported that the introduction of lithium atoms opens a band gap [7], turning Gr into a semiconductor material suitable for electronic applications; titanium-doped Gr is instead reported to enhance the gravimetric density for hydrogen storage applications [8], while doping of Gr with boron and nitrogen (N) atoms results in a higher electrical conductivity [9] and enhances its electrocatalytic activity towards oxygen reduction reaction (ORR) [10-12] and its potential for energy storage [13-15]. In particular, nitrogen doped Gr (N-Gr) is predicted to have a strong potential as a 2D material for several applications, including, for example, gas sensing and gas storage [16-18].

Unfortunately, the production of N-Gr is not trivial. Different synthesis methods have been proposed, following two main approaches: (i) a direct synthesis, in which N dopants are introduced during the Gr growth, and (ii) a post-synthesis approach, in which the Gr layer is doped after its growth. Among the direct synthesis methods, the most common is chemical vapor deposition (CVD) [19-21], in which a catalytic substrate is exposed to N- and C-containing precursors that dissociate on the surface, leading to the formation of high-quality N-Gr layer. However, the high pressure and temperature required for the growth process, the specific equipment and the dangerous precursors needed, make the production troublesome and pose serious safety issues. On the other hand, among the post-synthesis approaches, the most commonly used method is the nitrogen plasma bombardment of a previously grown Gr layer, promoting the replacement of some C atoms with N atoms [22, 23]. However, this procedure drastically reduces the quality of the Gr layer, due to the large defects induced by N bombardment.[24]

An alternative approach, taking advantage of N segregation from the substrate in order to produce N-Gr, was proposed in the literature [25]. The authors engineered a system based on a $SiO_2$/Si substrate, on which a boron layer, where a trace amount of nitrogen species was spontaneously incorporated, was firstly deposited by e-beam evaporation; a nickel (Ni) film was then deposited above it. Upon vacuum annealing, boron-trapped nitrogen and nickel-trapped carbon impurities diffuse towards the nickel surface and form N-Gr, while boron atoms remain deeper in the layered substrate. The authors suggested therefore that boron film acts as a nitrogen source carrier, while maintaining the presence of boron atoms into the graphene below their detectable threshold.

Starting from that concept, we present here an alternative, cleaner and highly reproducible growth method, based on a simpler and more controlled preparation procedure, which ensures the formation of layers of high morphological quality and is potentially easily scalable, from small samples prepared under ultra-high vacuum (UHV) conditions to large foils produced in industrial lines. The resulting N-Gr layers were thoroughly characterized by Scanning Tunneling Microscopy (STM) and X-ray Photoelectron Spectroscopy (XPS), in combination with an extensive Density Functional Theory (DFT) investigation, including simulated STM images and XPS spectra for a wide set of models for the N centers, to unveil their structural, morphological and chemical properties.

**2. Material and methods**

2.1 STM and XPS Experiments:

N-doped Gr layers were prepared in a UHV chamber with a base pressure of ~2 ×$10^{-10}$ mbar. A Ni(111) single crystal was cleaned by several cycles of $Ar^+$ sputtering at 1.5 kV at room temperature (RT) and annealing at 973 K, for a few minutes. Standard Gr growth was performed in UHV by low-pressure CVD, using ethylene ($C_2H_4$) as precursor. Low energy electron diffraction (LEED) and STM characterization was performed in UHV in order to assess the quality and homogeneity of the as-grown Gr sample. STM measurements were performed at room and cryogenic temperature (77 K) with an Omicron variable-temperature (VT) STM and an Omicron low-temperature (LT) STM. All topographic images were acquired in constant-current mode. All dI/dV conductance maps were acquired in constant-height mode using the lock-in technique. STM images were analyzed with the Gwyddion software package [26] and moderate noise filtering has been applied. Crystallographic orientation of the images was obtained by analyzing the epitaxial structure formed by pristine Gr on the Ni(111) surface, as described in Ref [27]. XPS measurements were performed *ex-situ*, with laboratory X-Ray sources and a base pressure in the range of $10^{-10}$ mbar, at two beamlines of the Elettra facility in Trieste (APE and ANCHOR-SUNDYN[28]) and at the IOM-INSPECT laboratory. All the spectra were collected at RT in normal emission geometry using different experimental apparatuses, X-ray sources (hν = 1486.7 eV, and hν = 1253.6 eV for the Elettra and IOM apparatuses, respectively) and hemispherical electron energy analysers. All binding energies were calibrated by measuring the Fermi level. The best fit of the experimental XPS data in the N1s region, was obtained using a Shirley background and 3 Doniach-Sunjic peaks, with binding energies converging to 397.1±0.15eV, 398.6±0.15eV and 400.6±0.15eV. A complex picture regarding the binding energies of N defects in a graphene overlayer exists in the literature: the pyridinic defects are found in the 397.8 eV - 401.8 eV binding energy range (depending on the N coordination and the possible presence of adsorbates), while the graphitic defects are found in 400.2 eV - 401.8 eV energy range [21, 25, 42, 43]. Pyrrolic defects, which are reported not to be stable at high temperature (~773 K) are generally found in the energy range 400.1 eV - 400.5 eV. [21]

2.2 DFT calculations:

Density Functional Theory (DFT) calculations were performed using the plane-wave-based Quantum ESPRESSO package (QE) [29, 30]. Ultrasoft pseudopotentials [31] were adopted to describe the electron-ion interaction with Ni (3d, 4s), C (2s, 2p), N (2s, 2p), treated as valence electrons. Energy cutoffs of 30 Ry and 240 Ry (for kinetic energy and charge density expansion, respectively) were adopted for all calculations. The Perdew-Burke-Ernzerhof functional (PBE) was used for electron exchange-correlation [32]. In order to properly describe the Gr/Ni(111) interaction, semiempirical corrections accounting for the van der Waals interactions were included with the DFT-D2 formalism [33]. Spin polarization was always included. For the simulation of Gr/Ni(111) interfaces, a 6 × 6 supercell with a total of 108 Ni atoms and 72 Atoms in the Gr layer was used, with a 2 × 2 × 1 Monkhorst-Pack k-points mesh [34] and vacuum of about 15 Å in the direction perpendicular to the surface to avoid interaction between images. The Ni(111) surface was modeled by a three-layer slab with a bottom layer fixed to the bulk positions during the geometry relaxation to mimic a semi-infinite solid. STM simulations were performed using the Tersoff-Hamann approach [35], according to which the tunnelling current is proportional to the energy-integrated Local Density of States (ILDOS). Constant-current and voltage values for the STM simulations have been chosen to match the experimental values. Ball-and-stick models and STM images were rendered with VESTA [36] and Gwyddion [26] softwares, respectively. For further details, see our previous work on N-doped or defective graphene on other metal substrates [37,38]. The XPS spectra of defective N-doped Gr have been calculated through the ΔSCF approach [39], which includes a pseudopotential generated with a full core hole for each ionized inequivalent atom and allows the calculation of relative changes of binding energies, called core level shifts (CLS). In order to compare the CLSs of different defects, which are inserted into separated supercells, we have included into each system a single $N_2$ molecule, at a distance of 16 Å from the surface, minimizing their mutual interaction. Such an approach has already been successfully employed for the calculation of the XPS spectra of several N-containing molecules, for example as done in Ref. [40], although with a different exchange-correlation functional. However, we have validated this method with respect to a direct comparison using two free-standing models of graphitic and pyridinic single nitrogen defect on Gr, respectively: the same separation of ~ 4.1 eV between their CLSs has been calculated (see Table S2 and the associated caption in the Supporting Information for a more detailed discussion).

## 3. Results and Discussion

The Ni(111) crystal was first exposed to a direct flux of porphyrins for few minutes at a temperature around 723 K in UHV. The chosen temperature conditions and the catalytic nature of Ni promote the cracking of the molecules on the metal surface and, in synergy with the high N solubility in the Ni crystal (comparable to that one of C) [41], lead to the creation of a nitrogen reservoir in the metal bulk (see Figure 1 and Figure S1).

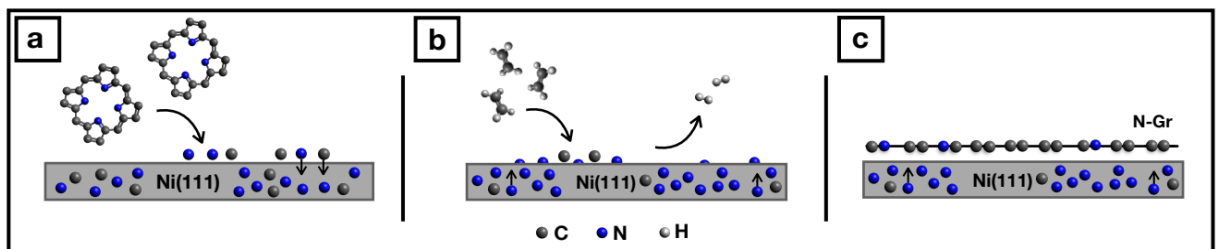

**Figure 1.** Schematic illustration representing: (a) Ni doping process with N atoms; (b) subsequent exposure to ethylene in UHV conditions; (c) formation of N-Gr.

The standard CVD Gr growth process was then started: the Ni(111) substrate, kept at 853 K, was exposed to ethylene ($C_2H_4$, partial pressure $5\times10^{-7}$ mbar) for 1 h. This leads to the formation of a complete layer, as for pristine Gr [27]. Finally, the sample was kept at the growth temperature for 10 more minutes, and cooled down to RT at a rate of 2 K s$^{-1}$. In this way, some of the N atoms segregating to the surface were trapped into the growing carbon network, leading to the formation of N-Gr.

We first investigated the chemical composition of the prepared sample by XPS, to verify the absence of contaminants and the presence of N in the 2D layer. Survey spectra ruled out the presence of any heterospecies except for the N introduced via exposure and cracking of porphyrin molecules. Figure 2a presents the typical XPS spectrum measured in the N 1$s$ region after several CVD cycles after nitrogen doping; the inset at the top left shows the spectrum acquired after just few cycles. The presence of several peak components indicates the co-existence of different chemical N configurations. Based on data fitting and on literature binding energies (see 2.1), it is possible to distinguish different N-containing species, with concentrations changing with CVD cycling. At the beginning (see inset), N is present in the form of (i) nickel nitride (397.1 eV), stemming from N atoms in the very first layers of the Ni substrate in a Ni nitride-like coordination, (ii) pyridinic defects (398.6 eV), where N is placed at the edge of a C vacancy and bonds to two C atoms as part of a six-membered ring, and (iii) graphitic defects (400.6 eV), where N atom is substituting a C atom in the Gr network and bonds to three neighbouring carbon atoms (see Figure 2a).

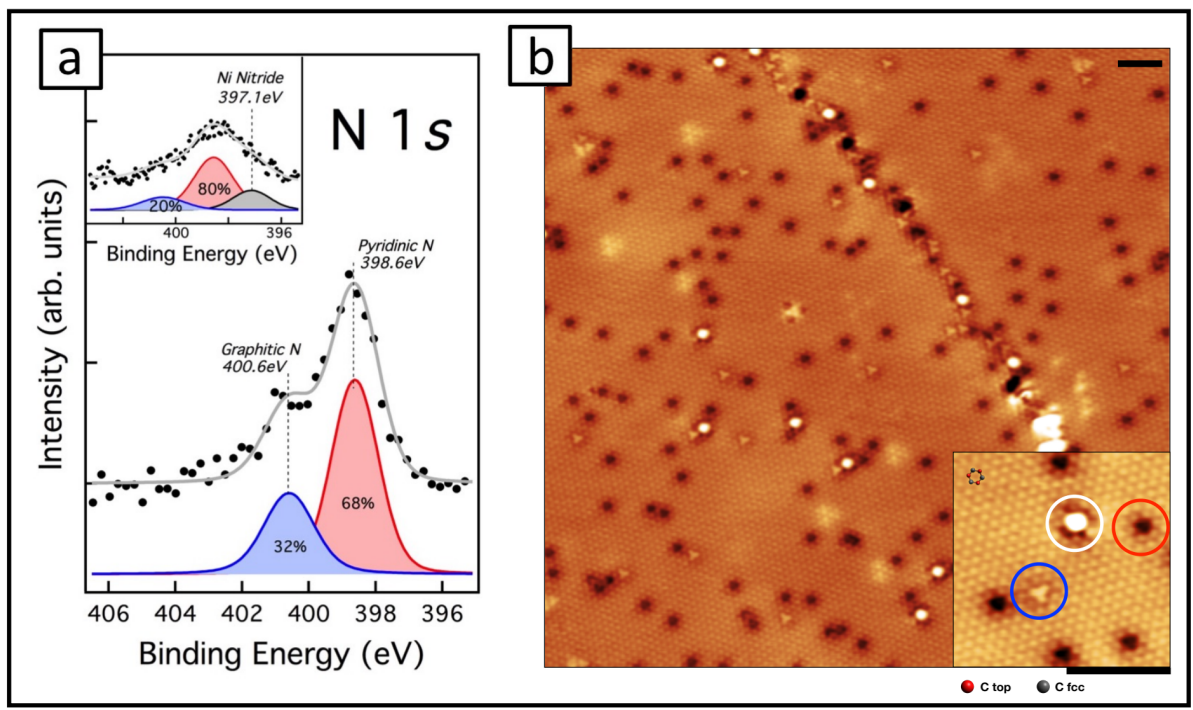

**Figure 2.** (a) XPS spectrum of the N 1*s* core level, after few (inset) and several (main spectrum) CVD cycles. Different N-containing species are highlighted. Photon energy hν = 1486.7 eV. (b) LT-STM image acquired at 77 K. Several defects are present on the surface. In the inset, the three most abundant types of defects are visualized: for explanation see main text. Scale bars 2 nm, I = 2.0 nA, V = -0.2 V.

The latter two components are a clear indication that N atoms are not exclusively stored in the Ni substrate, as evidenced by the lower energy component, but also trapped in the Gr mesh. After several CVD cycles, the nitride-like component is not visible anymore (see main spectrum in panel a), suggesting that the procedure has depleted the N content in the first Ni layers, while the fingerprint of N-doping of Gr is still evident. The XPS assignment above is confirmed by DFT simulations: the spectral separation calculated between the graphitic and pyridinic defects in Gr on Ni(111) is about 2.1 eV, in very good agreement with the experimental value of 2.0 eV. This value of the chemical shift is obtained for the most abundant graphitic and pyridinic defects observed by STM (see text below). For an extended comparison between the relative XPS binding energies calculated for different N defects and configurations see Table S1 in the Supporting Information.

The typical LEED pattern of the epitaxial Gr phase on Ni(111) (in Figure S2) and the experimental STM images (in Figure 2b) show that a Gr layer of good morphological quality was obtained. The surface appears flat and regular, as normally observed for Gr grown epitaxially on the Ni (111) substrate. A grain boundary is visible, separating two regions that have the same top-fcc geometry [44]. This confirms that the Gr layer was formed on the surface starting from different seeds that, during the growth, have given rise to distinct Gr domains. This is totally analogous to the case of pristine Gr, thus indicating that the N doping method presented in this work does not affect significantly the standard epitaxial Gr growth process. In particular, it is likely that carbon atoms coming from cracked porphyrins simply add to the bulk-dissolved carbon and then contribute to the formation of the layer as reported for un-doped graphene [27]. The Gr surface appears sprinkled by a huge number of defects with diverse appearance. The large white spots, clearly visible in Figure 2b and in its inset (white circle), are also observed in the pristine case (see Figure S3) and have been previously assigned to Ni atoms trapped in the Gr network during the growth process [45,46]. Indeed, it has been experimentally and theoretically demonstrated that Ni surface adatoms actively participate in the CVD Gr growth, catalyzing the C-C bond formation [45], remaining at times trapped into the Gr network [27]. Besides the trapped Ni adatoms, new kinds of defects were observed in the high-resolution images. Since these defects were not present in the pristine Gr case (see Figure S3), they can be tentatively linked to N atoms trapped in the Gr mesh. The inset reported in Figure 2b shows in detail two of the most abundant features of this kind, appearing as clover-like defects (blue circle) and dark triangles (red circles).

Atomically resolved images of all the observed N related defects are reported in the top row of Figure 3. For the defects in panels (3a) and (3b), the Gr network appears to be complete, with no vacancies in the mesh. This points to the presence of N atoms fully embedded in the Gr network, where they replace C atoms, in a graphitic configuration, in line with the XPS results discussed above.

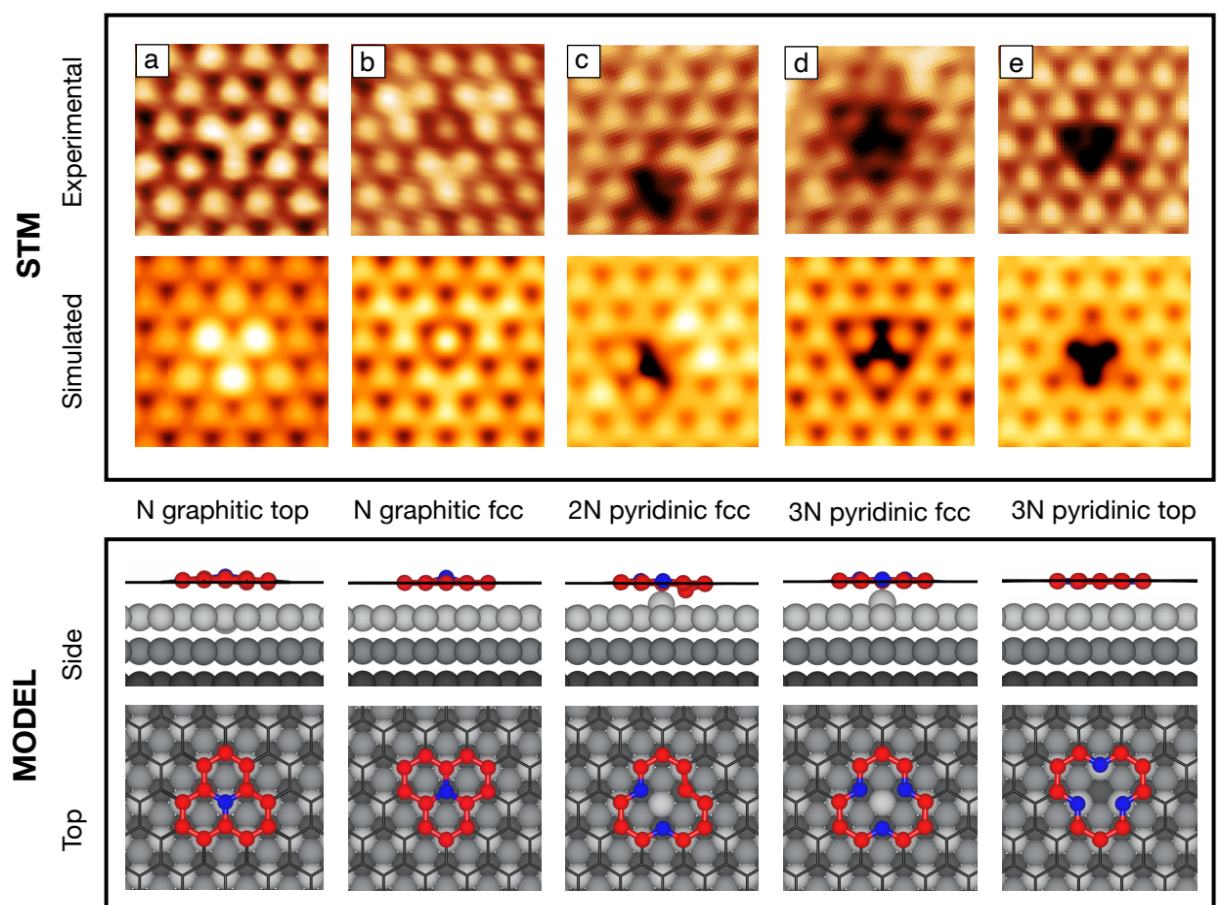

**Figure 3.** N defect configurations in Gr on Ni(111). Top panel: Experimental and simulated STM images for various defects. Image size: 1×1 nm$^2$. Structures are classified according to number of N atoms, configuration type and position. Experimental parameters: (3a) N graphitic top [I = 3 nA, V = -0.2 V], (3b) N graphitic fcc [I = 4 nA, V = -0.2 V], (3c) 2N pyridinic fcc [I = 0.6 nA, V = -0.2 V], (3d) 3N pyridinic fcc [I = 0.6 nA, V = -0.2 V] and (3e) 3N pyridinic top [I = 0.6 nA, V = -0.2 V]. Computational parameters: V = - 0.2 V; ILDOS isosurface lying ≈ 2 Å above Gr and with ILDOS value of $5 \times 10^{-5}$ |e|/$a_0^3$. Bottom panel: ball-and-stick model of DFT relaxed structures (side and top view). Color coding: Ni atoms in the first, second and third layer are rendered in dark grey, grey and light grey, respectively; N atoms in blue; C atoms delimiting the defect site in red, Gr network in black.

The clover-like defect (Figure 3a) is centered in a top site, with the first fcc neighbouring C atoms appearing brighter than the other C atoms of the mesh, which suggests a localized increase in the density of states. On the other hand, for the defect centered in an fcc position (Figure 3b), a localized intensity decrease is present at the position of the first neighbours of the central atom, accompanied by the appearance of brighter clovers formed by C atoms in symmetric positions. Considering their registry, we assign these two defects to top (3a) and fcc (3b) graphitic N configurations, respectively. The identification is confirmed by DFT calculations and simulated STM images of the proposed models, shown in the bottom panels, which nicely reproduce the appearance of these defects.

At variance with Figure 3a and 3b, Figure 3c, 3d and 3e show images characterized by a dark core, suggesting that the corresponding defects involve one or more atomic vacancies in the network. More specifically, the defect imaged in Figure 3c displays a mirror symmetry, whereas those in Figure 3d and 3e show a threefold symmetric shape, centered in fcc and top positions, respectively. In Figure 3d, the down-pointing black triangle shows three protrusions in fcc position close to the corners, with a slightly less bright appearance than C atoms in the network; conversely, the triangular defect in Figure 3e displays a completely black core. Due to the likely presence of vacancies, it is reasonable to expect that these three defects involve nitrogen in pyridinic structures; however, to rule out other possibilities, we did not limit our models to such configurations. In particular, for the defects in Figure 3c and in Figure 3e we have considered a wide variety of possible models (involving different numbers of vacancies and N atoms and considering also possible Ni adatoms, as shown in Figure S4). The comparison between the experimental STM images and those obtained from the extensive DFT investigation lead to the following assignment (see bottom panels): the defect in Figure 3c is compatible with a 2N pyridinic fcc with mirror symmetry, whereas the ones in Figure 3d and in Figure 3e can be assigned to three-fold symmetric pyridinic configurations of three N species in fcc and top positions, respectively. These 3N pyridinic models consist of three N atoms, each bound to two C atoms, at the edge of a C vacancy. The presence of pyridinic species in the N-Gr layer is fully consistent with the XPS measurements reported and discussed above. In particular, the 3N pyridinic top configuration is the one that best fits the experimentally observed XPS shift with respect to the graphitic top species (calc. 2.1 eV), as mentioned before and detailed in Table S1 in the Supporting Information.

A direct quantitative comparison between the amounts of 3N pyridinic and graphitic defects measured by XPS and STM is not trivial, as we could not perform *in situ* XPS measurements on the same layers imaged by STM. In any case, both techniques agree in indicating that the amount of N atoms in pyridinic configuration largely exceeds that in graphitic configurations. However, STM on average estimates the presence of more than twice pyridinic N than XPS. The reason of this discrepancy remains unclear and might be related to the influence of the different history of the measured layers and/or to effects such as photoelectron diffraction on the spectroscopic measurements, whose investigation is outside the scope of the present paper.

The following two facts are worth noting: 2N pyridinic fcc defects are very rare and their identification is not straightforward, while pyridinic configurations with 2N top, 1N top and 1N fcc, are never observed in our experimental STM images of the N-Gr layer. In order to understand why these structures are not present, we performed a set of DFT calculations aimed at evaluating the energy cost/gain of having N atoms at either pyridinic or graphitic positions in the lattice. We compared the total energy of three model configurations with the same number of N atoms (3N) but different relative positions (see Figure 4).

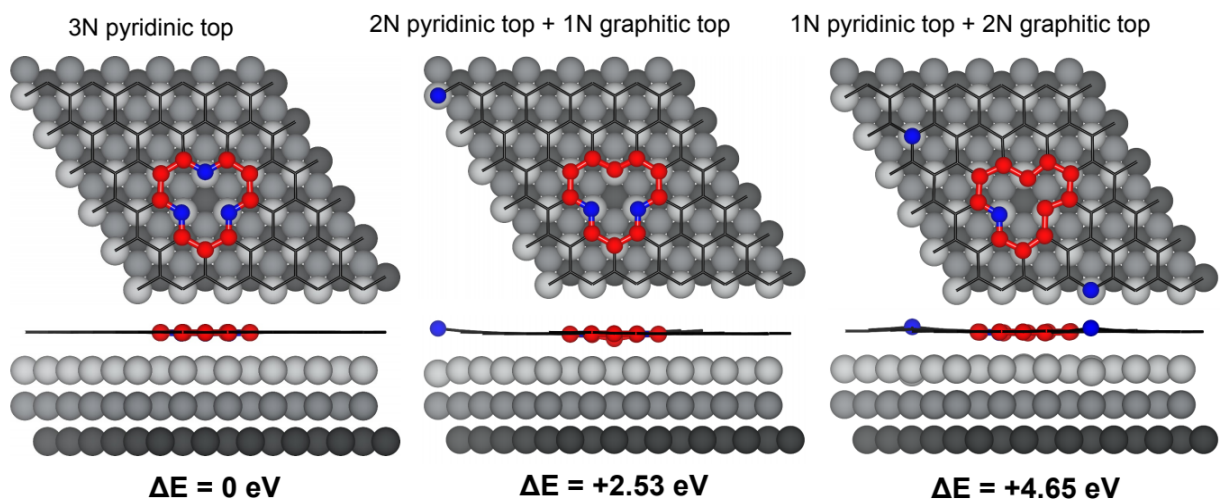

**Figure 4.** Top and side views of three different N-Gr models involving 3 Nitrogen atoms, classified according to the configuration type and position of the N atoms. The relative energy with respect to the most stable model (3N pyridinic top) is reported below each configuration. Color coding: Ni atoms in the first, second and third layer are rendered in dark grey, grey and light grey, respectively; N atoms in blue; C atoms delimiting the defect site in red, Gr network in black.

The most stable model (3N pyridinic top) is taken as energy reference. Both the other tested structures ("2N pyridinic top+1N graphitic top" and "1N pyridinic top+2N graphitic top") are characterized by higher total energies, +2.5 eV and +4.6 eV, respectively. This is a clear indication of the preference for the N atoms to be located at the edges of the C vacancy site rather than elsewhere, replacing a C atom within the Gr matrix, which explains the experimentally observed higher abundance of pyridinic defects with respect to graphitic ones. Such balance is likely related to the doping technique: adding N atoms during growth conditions favours the dynamic selection of defects with lower energy, with no constraint posed by the local abundance of N atoms required. Conversely, other doping techniques such as post-growth ion implantation favour defects with a different graphitic/pyridinic ratio, as a consequence of the ballistic formation process [21, 47, 48].

We further characterize the electronic structure of the observed N defects by acquiring STM images at different bias voltages. This allows the empty and filled states near the Fermi level to be probed. The comparison between the experimental and the simulated STM images shows a good agreement for N graphitic top and fcc defects, as well as for the 3N pyridinic fcc defect (see Figure S5). Only in the case of the 3N pyridinic top defect, the simulated image did not adequately reproduce the experimental features for positive bias polarity. Indeed, for this case simulated STM images show a dark core at all the considered biases (negative and positive), while the experimental images display a dark core only at negative biases and bright protrusions at positive ones. We do not have a clear explanation for this effect; it can be tentatively related to an enhancement of the inelastic scattering of tunneling electrons due to vibrational excitations of possible adsorbates on the tip [49], when the tip is located on specific defects, such as the 3N pyridinic with an fcc vacancy. To further corroborate our assignment based on the good agreement between experimental and calculated XPS shifts and STM images at negative bias, we have performed STS measurements at cryogenic temperature (77 K) to compare the spatial distribution of electron tunneling at specific energies with the localized density of states [50]. Indeed, experimental dI/dV conductance maps of graphitic defects, acquired at several biases, nicely match the corresponding maps of LDOS, (see Figure S6), providing further support to our interpretation about the nature of the observed N defects.

It should be noted that in all the above analysis we never considered pyrrolic-N defects. This is due to the fact that in our STM images we never observed defects of pentagonal shape, as expected for pyrrolic-N (one N atom in a 5-member ring); on the other hand, we could successfully reproduce the appearance of all the main N-defects observed using graphitic and pyridinic configurations. Furthermore, the XPS fitting results do not improve significantly if we include a further pyrrolic component. On this basis we can safely assume that, although we cannot exclude the presence of some pyrrolic-N defects, their coverage is not relevant.

Finally, STM images acquired at different temperatures and/or after annealing of the sample (see Figure S7) show no significant differences in the appearance of the defects, indicating that they are stable in a wide temperature range (from 77 K up to 520 K).

## 4. Conclusions

In summary, we showed that N-Gr can be formed on a Ni(111) substrate using an "inside out" growth method. N-Gr layers are obtained by dosing a commonly used hydrocarbon precursor ($C_2H_4$) on a hot Ni substrate, while N atoms segregate at the surface from a nitrogen reservoir previously created in the Ni bulk with a simple and reproducible doping process. The produced N-Gr layers were thoroughly characterized at the atomic level by STM and XPS in combination with DFT simulations. This approach yields continuous and flat layers, with a variety of N defects in the graphene mesh that we have identified and fully characterized. The formation of these defects is highly reproducible in different preparations using identical parameters and can potentially play a relevant role in tailoring unexpected electronic, optical and chemical properties of the graphene layer. To this purpose, it will be necessary to further explore the not straightforward influence of the specific parameters of the preparation procedure on the N defects present in the resulting layer, in order to achieve control of their distribution.

Considering the increasing interest of the scientific community in the improved properties of doped Gr, this work can contribute to open a new way to produce high-quality N-Gr layers with well-defined functional defects, suitable for new nanotechnological applications.

**Acknowledgements**

We acknowledge support from the Italian Ministry of Education, Universities and Research (MIUR) through the program PRIN 2017 - Project no. 2017NYPHN8. We thank Roberto